\documentclass[]{raa} 

\usepackage{aas_macros}
\usepackage{graphicx,times,color}
\usepackage{natbib}
\usepackage{amssymb,amsmath}
\bibpunct{(}{)}{;}{a}{}{,}

\usepackage{hyperref}
\hypersetup{colorlinks = true, linkcolor = green, anchorcolor = red, citecolor = blue, filecolor = red, pagecolor = red, urlcolor = red}

\newcommand{\diff}{{\rm d}}

\newcommand{\HI}{{H{\small\,I}}}

\begin{document}

\title{Blind Search for 21-cm Absorption Systems in New Generation Chinese Radio Telescopes}

\author{
Hao-Ran Yu\inst{1,2},
Ue-Li Pen\inst{2,3,4,5},
Tong-Jie Zhang\inst{6},
Di Li\inst{7,8} and
Xuelei Chen\inst{9}
}

\institute{
Kavli Institute for Astronomy \& Astrophysics, Peking University, Beijing 100871, China;\\ \and
Canadian Institute for Theoretical Astrophysics, University of Toronto, Toronto, M5S 3H8, ON, Canada;\\ \and
Dunlap Institute for Astronomy and Astrophysics, University of Toronto, Toronto, M5S 3H4, ON, Canada;\\ \and
Canadian Institute for Advanced Research, Program in Cosmology and Gravitation;\\ \and
Perimeter Institute for Theoretical Physics, Waterloo, ON, N2L 2Y5, Canada;\\ \and
Department of Astronomy, Beijing Normal University, Beijing, 100875, China;\\ \and
National Astronomical Observatories, Chinese Academy of Sciences, Beijing, China;\\ \and
Key Laboratory of Radio Astronomy, Chinese Academy of Sciences, Beijing, China;\\ \and
Key Laboratory for Computational Astrophysics, National Astronomical Observatories, Chinese Academy of Sciences, Beijing, 100012, China.\\
}

\abstract{
Neutral hydrogen clouds are known to exist in the Universe, however their spatial
distributions and physical properties are poorly understood. Such missing information
can be studied by the new generation Chinese radio telescopes through a blind searching
of 21-cm absorption systems. We forecast the capabilities of surveys of 21-cm absorption
systems by two representative radio telescopes in China --
Five-hundred-meter Aperture Spherical radio Telescope (FAST) and
Tianlai 21-cm cosmology experiment (Tianlai).
Facilitated by either the high sensitivity (FAST)
or the wide field of view (Tianlai) of these telescopes,
more than a thousand 21-cm absorption systems can be discovered in a few years,
representing orders of magnitude improvement over the cumulative discoveries
in the past half a century. 
\keywords{}
}

\authorrunning{Yu et al.}
\titlerunning{Blind Search for 21-cm Absorption Systems}

\maketitle

\section{Introduction}\label{sec.introduction}
Neutral hydrogen (\HI) clouds are known to exist in the Universe,
however relatively few of them are discovered in the past half a century,
and we poorly understand their spatial distribution and physical
properties \citep{2005ARA&A..43..861W}. 
In damped Lyman-$\alpha$ absorption (DLA)
systems, the radio spectrum is also substantially absorbed by the {\HI}
hyperfine structure, whose rest frame wavelength is approximately 21 cm
($1420405751.7667\pm 0.009$ Hz in frequency). These systems have
at least $2\times 10^{20}$ cm$^{-2}$ {\HI} column density, thus being
able to absorb the background with its cold {\HI} in their cold neutral medium (CNM).

The 21-cm absorption systems are important in the study of the 
distribution, location, temperature and structure of neutral gas,
and the evolution of neutral gas systems and galaxies over
cosmic time scale.
Due to the narrow intrinsic line width, the 21-cm absorption systems
are proposed to be used to directly measure the cosmic acceleration
\citep{2012ApJ...761L..26D,2014PhRvL.113d1303Y}, via the 
Sandage-Loeb (SL) effect \citep{1962ApJ...136..319S,1998ApJ...499L.111L}.
The main source of uncertainty of such proposed measurements lies in our
poor understandings of {\HI} clouds \citep{2014PhRvL.113d1303Y}.
New surveys are required
to improve our understandings of spatial distribution and physical
properties of {\HI} clouds.
This can be done by the Chinese new generation radio telescopes --
Five-hundred-meter Aperture Spherical radio Telescope (FAST)
\citep{2016RaSc...51.1060L} and Tianlai 21-cm cosmology experiment
(Tianlai) \citep{2012IJMPS..12..256C}.
The single-dish FAST, like Arecibo\footnote{https://www.naic.edu},
uses its gigantic single dish to achieve ultimate sensitivity, whereas
Tianlai, designed like CHIME\footnote{http://chime.phas.ubc.ca},
has ultra wide field of view (FoV) provided by its
using its cylindrical reflectors and arrays of receivers to
quickly scan the northern hemisphere of the sky as the Earth rotates.

Considering their respective design and observing strategies,
we forecast their capabilities of blind searching of 21-cm absorption
systems. In section \ref{sec.sensitivity} we show the sensitivity
estimation and related factors for FAST and Tianlai.
We present our forecast in section \ref{sec.FAST}
and section \ref{sec.Tianlai} respectively.
Conclusions are made in section \ref{sec.conclusion}.

\section{Sensitivity}\label{sec.sensitivity}
To find possible 21-cm absorption systems, radio telescopes
could be devised to scan the radio sources in
NRAO VLA Sky Survey (NVSS)\footnote{http://www.cv.nrao.edu/nvss/}.
If we assume the distributions of radio sources and
{\HI} clouds are uncorrelated over the sky, the redshift
distribution of the 21-cm absorption systems $n_{\rm system}$ would be
\begin{equation}\label{eq.n_system}
    n_{\rm system}(z)=n_{\rm \HI}^*(z)\int_z^\infty n_R(z')\diff z',
\end{equation}
where $n_R$ is the redshift distribution of radio sources over the sky
(given by equation (26) of \cite{2010A&ARv..18....1D}), and
$n_{\rm \HI}^*(z)$ is the number of occupation of {\HI} clouds
per any given line of sight. Note that, unlike $n_R$ and $n_{\rm system}$,
 $n_{\rm \HI}^*(z)$ is not an
integration of volume so we do not need a volume filling factor in
equation \ref{eq.n_system}.
However, this spacial distribution of {\HI} is poorly understood and is to be
studied in upcoming surveys. Recent studies show that the number density of
absorbers to be $\diff n_{\rm \HI}/\diff z=n_{\rm \HI}^*(z)= 0.045\pm0.006$
per line of sight at low and medium redshift
\citep{2005ARA&A..43..861W,2007ASSP....3..501Z}, so we apply this value
in the following forcast. Equation \ref{eq.n_system} gives the
{\HI} cloud occupation regardless of observabilty, based on which
we calculate the detections considering sensitivities.

Observation sensitivity depends on the background fluxes and foreground
absorptions (current section), as well as telescope configurations and
observation strategies (section \ref{sec.FAST},\ref{sec.Tianlai}).

NVSS contains 2 million sources (with declination $\delta >-40^\circ$,
cover 82\% of the sky) stronger than 2.5
mJy at $\nu=$1.4 GHz \citep{1998AJ....115.1693C} and their flux
distribution $n_R(F)$ is given by \cite{1984ApJ...287..461C}.
The observed flux is already lowered by the 21cm absorption if there is any on
the line of sight, so the actual signal to noise of the absorption line
is higher. Due to the limited knowledge of absorption system properties
we ignore this factor for now.
Redshifted {\HI} clouds absorb lower frequency bands where
the radio sources are typically brighter by $\nu^{-0.7}$
\citep{1998AJ....115.1693C}, or $(1+z_{\rm \HI})^{0.7}$.
The error of the measurement $\Delta F$ is given by\footnote{https://science.nrao.edu/facilities/vla/docs/manuals/oss/performance/sensitivity}
\begin{equation}\label{eq.delta_F}
    \Delta F=T_{\rm sys}A_{\rm eff}^{-1}/\sqrt{n_{\rm pol}\Delta\nu\Delta t},
\end{equation}
where
$T_{\rm sys}$ and $A_{\rm eff}$ are system temperature and effective receiving area,
and $T_{\rm sys}A_{\rm eff}^{-1}$ is just the system equivalent flux density (SEFD).
%SEFD \tcr{(or $T_{\rm sys}$ TBD)} is the system equivalent flux density,
$n_{\rm pol}$
is the number of polarizations, $\Delta\nu$ is the line width and
$\Delta t$ is the integration time. $\Delta\nu=(u_{\rm width}/c)\nu_{\rm obs}$,
where $u_{\rm width}$ is the equivalent line width of the {\HI}
absorber. The properties of 21-cm absorbers are primarily derived from
followup studies of optical absorbers. The discovery rate from the
cross correlation is a lower bound on the expect number of absorbers,
since high column density systems in the CNM may systematically
obscure potential background optical sources \citep{2014PhRvL.113d1303Y}. There are
only 3 blind radio detections, and a survey may discover more systems
which are optically obscured. They would likely be cold and at high
column density. For sensitivity purposes, we treat all sources as
$u_{\rm width}=2\ {\rm km\ s}^{-1}$
\citep{1982ApJ...259..495W,2005ARA&A..43..861W}.
$\nu_{\rm obs}$ is the observation frequency. For redshifted
21-cm absorption systems, $\nu_{\rm obs}=1420\ {\rm MHz}/(1+z_{\rm \HI})$.
The integration time $\Delta t=(n_{\rm obs}\times 24\ {\rm h})\tau$,
where $n_{\rm obs}$ is the number of days an object is scanned,
and $\tau=\lambda_{\rm obs}/2\pi D\cos\delta$ is the fractional time the object
transits the FoV ($\lambda_{\rm obs}\ll D$ or $\delta\nrightarrow\pi/2$).
For 21-cm absorption systems $\lambda_{\rm obs}=21\ {\rm cm}\times (1+z_{\rm obs})$.
$\Delta F$ is redshift- (frequency-) independent, as $(1+z_{\rm obs})$ terms
from integration time and line width cancel out
in the equation (\ref{eq.delta_F}).

In the forecast we count $>10\sigma$ detections, i.e., the absorption
line depth is at least $10\Delta F$. The optical depth of {\HI} absorbers
are also poorly understood. Recent discoveries of 21-cm absorption systems
have about $r=20\%$ fractional depth \citep{2015MNRAS.453.1249A,2015MNRAS.453.1268Z},
and we apply this value. Thus, all systems with $rF>10\Delta F$ are counted.
Note that there is no confusion in the spectral domain -- 
When the system temperature is not dominated by sources,
the noise in each pixel is not affected by the number of sources.
So only sources with absorbers contribute to the signal, 
and nothing contributes to the noise.
Thus being confusion limited makes no difference.

\section{FAST estimation}\label{sec.FAST}

FAST has a primary dish of 500 meter in diameter. The effective
antenna aperture of FAST is 300 m, in diameter for any zenith angle
up to $26^\circ$ and decreases to 200 m for a maximum zenith angle 
of $40^\circ$ \citep{2016RaSc...51.1060L}.
%This enables its zenith angle up to $40^\circ$ and it can scan objects
%in declination range $-14^\circ 12'<\delta<65^\circ 48'$.
The 19-beam feed-horn array at L band will be the primary survey
instrument for FAST. For a single day it can scan total of $0.5^\circ$
in declination $\delta$. FAST survey strategy scans each object
only once ($n_{\rm obs}=1$), and with its high sensitivity,
even faint sources are identified in a single day. 
We assume a one-month survey by FAST
around the celestial equator for simplicity. FAST observes
1.02 to 1.42 GHz (redshift $z_{\rm \HI}<0.39$), so objects on celestial equator
have average transit time $\Delta t \simeq 12\ {\rm sec.}$
Taking $\Delta\nu\simeq$ 9 kHz, $n_{\rm pol}=2$ (dual polarized system),
%\tcr{$T_{\rm sys}\sim 20$ K and $A_{\rm eff}\sim 7\times 10^4$ m$^2$,
%we get ${\rm SEFD}\sim 1$ Jy for FAST},
FAST has SEFD of $\sim 1$ Jy,
then we have $\Delta F \simeq\ 2.3\ {\rm mJy}$.
Taking into account the $\nu^{-0.7}$ flux boost at lower
frequencies, $rF>10\Delta F$ requires $F\gtrsim 0.11\ {\rm Jy}$ on 1.4 GHz.

One-month-scan ($n_{\rm obs}=30$) around celestial equator by FAST will have
$N_R\simeq 10^4$ sources whose $F\gtrsim 0.11\ {\rm Jy}$ on 1.4 GHz.
We further assume the independency between $n_R(F)$, $n_R(z)$
and $n_{\rm\HI}(z)$, from $n_{\rm system}(z)$ by integrating
equation (\ref{eq.n_system}), there will be $N_{\rm system}\simeq 200 {\rm /month}$ absorption
systems found in redshift range $0<z<0.39$.

\section{Tianlai estimation}\label{sec.Tianlai}

Tianlai is currently $30\times 12\ {\rm m}$ area with
%\tcr{${\rm SEFD}=T_{\rm sys}A_{\rm eff}^{-1}\simeq 300\ {\rm Jy}$}. 
${\rm SEFD}\simeq 300\ {\rm Jy}$.
Different from FAST, however,
Tianlai scans all sources in the northern hemisphere of the sky
everyday, and one needs longer periods integration to
identify absorbers with fainter background sources.
We forecast its one-year-survey capability of searching 21-cm
absorption systems in the northern hemisphere.

Because Tianlai locates at latitude
$\phi_{\rm site}=+44^\circ$ and its cylinders are fixed, only
objects at the zenith ($\delta=\phi_{\rm cite}=44^\circ$) fully utilize
$A_{\rm eff}$. On the other hand, higher declination objects
have longer integration time per day. Thus, $\Delta F$ is
a function of $\delta$, 
\begin{equation}\label{eq.delta_F_delta}
    \Delta F(\delta) = \frac{T_{\rm sys}A_{\rm eff}^{-1}}{\cos(\delta-\phi_{\rm site})}
    {\sqrt{\frac{\cos\delta}{n_{\rm pol}\Delta\nu\Delta t(\delta=0)}}},
\end{equation}
where $\Delta t(\delta=0)\simeq 1.17\times 10^5\ {\rm sec}$ is the
total integration time (1-year survey) for objects on celestial equator,
for Tianlai frequency range 800 to 900 MHz ($0.58<z<0.78$).
The effective $N_R$ is given by
\begin{equation}\label{eq.N_R_eff}
    N_R=\int_0^{\pi/2}2\pi\cos\delta
    \left(\int_{10\Delta F(\delta)/r}^{+\infty}
    n_R(F')\diff F'\right)
    \diff\delta.
\end{equation}
Note that the inner flux integration converges as
$F'\rightarrow+\infty$ because $n_R(F)\sim F^{-2.5}$.
Additionally, even if $\tau=\lambda_{\rm obs}/2\pi D\cos\delta$
fails at $\delta\rightarrow\pi/2$ (the north pole
$\tau(\delta=\pi/2)=1$ is always in FoV),
the outer declination integration also converges
as $\delta\rightarrow\pi/2$, because the pole area is
tiny and neglectable.
For FAST, surveys at higher declinations would as well
use equation (\ref{eq.delta_F_delta},\ref{eq.N_R_eff}),
however $\cos(\delta-\phi_{\rm site})$ term affecting
SEFD should be neglected.

For Tianlai, equation (\ref{eq.N_R_eff}) gives $N_R\simeq 1.3\times 10^4$,
and applying it to equation (\ref{eq.n_system}) again
we get $N_{\rm system}\simeq 80 {\rm /year}$ absorption
systems found in redshift range $0.58<z<0.78$.

\section{Conclusion}\label{sec.conclusion}
We forecast the capability of FAST and Tianlai telescopes to search 21-cm
absorption systems. According to our assumptions, FAST is able
to find $\simeq 200$ systems per month
whereas TianLai can find $\simeq 80$ systems per year.
Comparing the results between two quite different telescopes --
high sensitivity FAST and large FoV Tianlai, we find the former is
more efficient in looking for fainter radio sources
($n_R(F)\sim F^{-2.5}$) for possible foreground absorptions.
Future improvements in the sensitivity on Tianlai enable it
more effectively looking for highest signal-to-noise systems
over a wider sky.

Regarding the quantitative forecasts in each telescope, although we take
into account many detailed aspects affecting the result, the major
uncertainties are $n_{\rm \HI}(z)$, $u_{\rm width}$ and $r$.
These poorly understood parameters are conversely worth investigating
from these proposed surveys. Modest real-time analysis changes could
allow the survey of 21-cm absorption systems. The data would need to be
recorded at sufficient spectral resolution. Spatial computational costs are
in principle unchanged, but there could be additional overhead
costs for the larger resulting data sets. Systematic surveys of
{\HI} clouds over a cosmic scale enable us to measure the 3D density
of these systems. For objects detected by FAST, one can also try to measure
the 21cm emission size. For Tianlai, one can try to measure the size from
very-long-baseline interferometry (VLBI).

\begin{acknowledgements}
This work was supported by the National Science Foundation of China (Grants No. 11573006, 11528306), the Ministry of Science and Technology National Basic Science program (project 973) under grant No. 2012CB821804.
HRY acknowledges General Financial Grant No.2015M570884 and
Special Financial Grant No. 2016T90009 from the China
Postdoctoral Science Foundation.
HRY and ULP acknowledge the support of the
National Science and Engineering Research Council of Canada.
\end{acknowledgements}

\bibliographystyle{raa}
\bibliography{haoran_ref}

\end{document}